\documentclass[useAMS,usenatbib]{mn2e}
\usepackage{graphicx} 
\usepackage{times} %michael reckons this gives more like journal font

\def\kms{km ${\rm s}^{-1}$}

\def\ch2{$\chi^2$}

\def\kms {\hbox{${\rm km\ s}^{-1}$}}

\def\scm  {$\hbox{{\rm cm}}^{-2}$}    %cm-2
\def\MOLH {\hbox{${\rm H}_2$}}  %H2
\def \AL {$\alpha $}     %  gr. alpha
\def \HI {H{\sc \,i}}

\def\lapp{\ifmmode\stackrel{<}{_{\sim}}\else$\stackrel{<}{_{\sim}}$\fi}
\def\gapp{\ifmmode\stackrel{>}{_{\sim}}\else$\stackrel{>}{_{\sim}}$\fi}
\def\bsp_small{\vspace{0.5cm}\small\noindent This paper has been typeset
from a \TeX/\LaTeX\ file prepared by the author.\normalsize}

\title[\HI\ 21-cm, Mg{\sc \,ii} and metallicities in
DLAs]{Relationships between the \HI\ 21-cm line strength, Mg{\sc \,ii}
equivalent width and metallicity in damped Lyman-{\boldmath $\alpha$}
absorption systems} 
\author[S. J. Curran et
al.]{S. J. Curran$^{1}$\thanks{E-mail: sjc@phys.unsw.edu.au},
P. Tzanavaris$^{2}$, Y. M. Pihlstr\"{o}m$^{3}$ and J. K. Webb$^{1}$ \\
$^{1}$School of Physics, University of New South Wales, Sydney NSW 2052, Australia\\
$^{2}$Institute of Astronomy and Astrophysics, National Observatory of Athens, I. Metaxa \& V. Paulou
152 36 Penteli, Greece\\
$^{3}$Department of Physics and Astronomy, The University of New Mexico, Albuquerque, NM 87131, USA\\
}
\begin{document}

\date{Accepted ---. Received ---; in original form ---}

\pagerange{\pageref{firstpage}--\pageref{lastpage}} \pubyear{2007}

\maketitle

\label{firstpage}

\begin{abstract}
We present the results of a survey for 21-cm absorption in four never
previously searched damped Lyman-$\alpha$ absorption systems (DLAs)
with the Westerbork Synthesis Radio Telescope. The one
detection is presented and discussed in \citet{ctm+07} and here we add
our results to other recent studies in order to address the important
issues regarding the detection of cold gas, through 21-cm absorption,
in DLAs: Although, due to the DLAs identified with spiral galaxies,
there is a mix of spin temperature/covering factor ratios at low
redshift, two recent high redshift end points \citep{kse+06,kcl06}
confirm that this ratio does not generally rise much above $T_{\rm
spin}/f\sim10^3$ K over the whole redshift range searched (up to
$z_{\rm abs}=3.39$). That is, if the covering factors of many of these
galaxies were a factor of $\geq2$ smaller than for the spirals (which
span $120\leq T_{\rm spin}/f\leq520$ K), then no significant
difference in the spin temperatures between these two classes would be
required.

Furthermore, although it is difficult to separate the relative
contributions of the spin temperature and covering factor, the new
results confirm that 21-cm detections tend to occur at low angular
diameter distances, where the coverage of a given absorption cross
section is maximised. This indicates a dominant contribution by the
covering factor. Indeed, the two new high redshift detections occur
towards two extremely compact radio sources ($\leq0.04''$), although
the one other new detection, which may have an impact parameter in
excess of $75$~kpc, occurs towards one of the largest radio sources
\citep{ctm+07}.

Finally, we also find an apparent 21-cm line strength--Mg{\sc \,ii}
equivalent width correlation, which appears to be due to a coupling of
the velocity structure between the components that each species
traces. That is, the gas seen in 21-cm absorption could be the same as
that seen in optical absorption. Combined with the known equivalent
width--metallicity relation, this may be manifest as a spin
temperature--metallicity anti-correlation, which is non-evolutionary
in origin.

\end{abstract}


\section{Introduction}\label{intro}

Although currently relatively rare\footnote{See table 3 of
\citet{cwm+06}.}, redshifted absorption systems lying along the
sight-lines to distant quasars are important probes of the early to
present day Universe. Of particular interest are damped Lyman-$\alpha$
absorption systems (DLAs), which contain at least 80\% of the neutral gas mass
density in the Universe \citep{phw05}. DLAs are believed to be the
precursors of modern-day galaxies and studies over a range of
redshifts are important to establish the link between the early stages
of galaxy formation and the galaxies known in detail today. However,
despite their importance in the context of galactic evolution, the
typical size and structure of DLAs has long been an issue of much
controversy, with models ranging from large, rapidly rotating
proto-disks (e.g. \citealt{pw97}) to small, merging sub-galactic
systems (e.g. \citealt{hsr98}). Moreover, imaging of DLA host galaxies
at $z\lapp1.6$ (where the galaxy can be distinguished against the
point spread function of the background QSO), reveals a mixed bag of spiral,
irregular, low surface brightness (LSB) and dwarf galaxies
(e.g. \citealt{bb91,lbbd97,cl03,rnt+03}). This variety being
confirmed by a blind 21-cm emission survey of local galaxies
\citep{rws03}.

Whatever their morphologies, the high abundance of cold neutral gas in
DLAs is expected to provide a reservoir for star formation at high
redshift. However, abundances of (the star forming) molecular gas in
DLAs are exceedingly low, \MOLH\ only being detected in a very few
cases
\citep{lv85,lmc+00a,lddm01,gbk01,lsp02,lps03,lps06,psl02,plns06,le03,rbql03,cbgm05,nlp+07},
which calls into question the ability of DLAs to contribute
significant star formation activity to the earlier Universe
\citep{lps03}.  Despite this, DLAs exhibit a (weak) evolution of
elemental abundance with redshift
\citep{pksh95,lsb+96,vbcm00,kf02,pgw+03}, which would be expected from
the enrichment of the interstellar medium in each galaxy by successive
generations of stars. Additionally, for the \MOLH-bearing DLAs, the
molecular fraction also exhibits an anti-correlation with redshift,
and the fact that the metallicity evolution is significantly steeper
for these DLAs may suggest that these actually constitute a narrower
class of objects (or sight-lines) than the general DLA population
\citep{cwmc03}. Furthermore, the decrease of molecular fraction with
look-back time indicates an evolution in the dust
abundance\footnote{Note that, at $z\sim3$, the general DLA population
shows no sign of reddening due to dust \citep{ml04} and that the
\MOLH-bearing DLAs have optical--near-IR colours of only $V-K=2.2 - 3.4$,
cf. $V-K\geq5.07$ for the five OH absorbers towards reddened quasars
\citep{cwm+06}.}, which is also evident through a decrease in the dust
depletion factor with redshift in the \MOLH-bearing DLAs
\citep{mcw04}.

Where the DLA occults a radio-loud QSO, the 21-cm transition can
provide complementary data to that obtained from the ultra-violet
Lyman-$\alpha$ observations (which are redshifted into the optical
band at $z\gapp1.7$). The former transition traces the cold gas,
whereas the latter traces all of the neutral gas, thus providing a
potential thermometer of the absorber. Provided that the 21-cm and Lyman-$\alpha$
absorption arise in the same cloud complexes, the spin temperature,
$T_{\rm spin}$ [K], of a homogeneous cloud can be derived from the
column density, $N_{\rm HI}$ [\scm], obtained from the Lyman-$\alpha$,
via
\begin{equation}
%N_{\rm HI}=-1.823\times10^{18}.T_{\rm spin}\int\!\ln\left(1-\frac {\sigma}{f.S}\right)\,dv\,,
N_{\rm HI}=1.823\times10^{18}\,T_{\rm spin}\int\!\tau\,dv\,,
\label{enew}
\end{equation}
where $\int\tau\,dv$ is the integrated optical depth of the 21-cm
line. However, obtaining the spin temperature from a 21-cm observation
is not quite so straightforward, since the observed optical depth of the 21-cm
line also depends upon on how effectively the background radio
continuum is covered by the absorber. Specifically,
$\tau\equiv-\ln\left(1-\frac{\sigma}{f\,S}\right)$, where $\sigma/S$
is the depth of the line relative to the flux density and $f$ is the
covering factor of the flux by the absorber.

Of the DLAs searched (and published) in 21-cm absorption, there is
currently a detection rate of slightly less than $50$\%. Most of these
occur at $z_{\rm abs}\lapp2$, where there is a roughly equal number of
non-detections. \citet{ck00,kc01a,kc02} attribute this distribution to
the low redshift DLAs having a mix of spin temperatures, whereas the
high redshift absorbers have exclusively high spin temperatures,
resulting in the large number of non-detections at high redshift (none
then detected at $z_{\rm abs}>2.04$). Since all of the DLAs identified
as large spirals have low spin temperatures, whereas those identified
with LSBs and dwarf galaxies, have higher spin
temperatures, \citet{kc02} conclude that the DLA host population
consists mainly of the warmer dwarfs and LSBs at high redshift, which
evolves to include a higher proportion of spirals at low
redshift. However, there are several caveats regarding this
conclusion:
\begin{enumerate}
	\item Although \citet{kc02} have undertaken a careful analysis
	in order to obtain the most likely covering factors, 
	estimating this from the flux of the compact unresolved
	component to the total flux of the quasar or by examining the quasar's
	spectral energy distribution, none of these methods provides
	any information on the size of the absorbing region.

	\item When such information is unavailable, more often than
	not in the literature the covering factor is usually assumed
	to have its maximum value of unity (notable exceptions are
	\citealt{br73,wbj81,bw83,wbt+85,cps92,lbs00,kc01a}, see table
	1 of \citealt{cmp+03}). If in reality $f<1$, this assumption
	would have the effect of assigning artificially high spin
	temperatures to DLAs. In the case of \citet{kc02}, half of the
	12 detections have $f<1$ and half $f=1$, whereas for the
	non-detections, 10 out of 11 have $f=1$, the vast majority of
	which (8) are at $z_{\rm abs}>1$ (see figure 1 of
	\citealt{cmp+03}).
	
	\item If the $T_{\rm spin}/{f}$ degeneracy is left intact,
	  there still remains a mix of values at low redshift, with
	  exclusively high values at high redshift (figure 5 of
	  \citealt{cmp+03}). However, of the non-detections only one
	  morphology is actually known (see Fig. \ref{Toverf}), and
	  although, due to their high ``spin temperatures'', it is
	  tempting to count these as non-spirals, this is not actually
	  known. In fact.  there are no host identifications at
	  $z_{\rm abs}\gapp0.9$.

	\item Furthermore, figure 3 \citet{kc02} suggests that at
	$z_{\rm abs}\gapp3$ spin temperatures in DLAs are expected to
	exceed $\sim2000$ K, with half of these $\gapp3000$ K,
	rendering these very difficult to detect. However, since then,
	21-cm absorption has been detected at $z_{\rm abs}=3.39$
	towards 0201+113 \citep{kcl06} at a spin
	temperature of $T_{\rm spin}\leq1950$ K (for $f\leq1$), when
	this was previously believed to have $T_{\rm spin}>3300$~K
	\citep{kc02}.
 
%\end{enumerate}

Additionally:
%\begin{enumerate}
	\item \citet{cmp+03} find that $\approx70$ per cent of the
	non-detections occult large background ($\gapp1$ arc-sec)
	sources and that 21-cm absorption tends to be detected
	towards these large sources only when the DLA host has been
	identified as a spiral.

\item Again, although for the non-detections the morphologies are
essentially unknown, lower covering factors, resulting in
non-detections at high redshift, would also be consistent with
hierarchical galaxy formation scenarios, where compact galaxies at
high redshift evolve to include a higher proportion of larger galaxies
in the more immediate Universe.

\item Most recently, \citet{cw06} find that all of the $z_{\rm
abs}\gapp1$ absorbers have large DLA-to-QSO angular diameter distance
ratios ($DA_{\rm DLA}/DA_{\rm QSO}\approx1$), whereas the $z_{\rm
abs}\lapp1$ absorbers have a mix of ratios. Thus reproducing the
``spin temperature'' segregation of \citet{kc02} through geometrical
effects alone. Since the DLAs with low ratios are almost always
detected in 21-cm absorption, this and the other points suggest that
the covering factor, rather than the spin temperature (which could
range from $170$ to $>9240$ K, \citealt{kc02}), is the important
criterion in determining whether 21-cm absorption is detected in a
DLA.
\end{enumerate}

We therefore expect 21-cm absorption to be readily detectable in DLAs
located along the sight-lines towards compact radio sources.  In order
to further address this, as well as finding new sources in which to
pursue our primary objective, the measurement of cosmological
variations in the fundamental constants at large look-back times
(e.g. \citealt{tmw+06}), we are undertaking a survey for \HI\ 21-cm
absorption in the suitably radio-illuminated DLAs yet to be searched:
In Section 2 we present the results of the first phase of this survey,
observations of both low and high redshift DLAs with the Westerbork
Synthesis Radio Telescope (WSRT). In Section 3 we discuss these and
the other new results (since \citealt{cmp+03}) in the context of
factors affecting the detectability of 21-cm in DLAs, as well as
investigating correlations between the 21-cm line strength, Mg{\sc
\,ii} equivalent width and metallicity in these absorbers.

\section{Observations and Results}\label{obs}

\subsection{Sample Selection}

The DLAs were selected from the Sloan Digital Sky Survey Damped
Lyman-{$\alpha$} Survey Data Release 1 \citep{ph04} and the known
systems also occulting radio-loud quasars ($S\gapp0.1$ Jy) as yet
unsearched \citep{cwbc01}\footnote{Available from
http://www.phys.unsw.edu.au/$\sim$sjc/dla}. Of those redshifted into
the WSRT's Ultra High Frequency (UHF) receiver bands, we selected an
initial sample which gave a mix of low and high redshifts in order to
test the arguments presented above. Since we require
deep observations in order to, at the very least, obtain useful
limits, each absorber was observed for 12.5 hours, which limited the
number of sources to four, although a further two candidate sub-DLAs were also
observed along the line-of-sight to 1402+044 (Table \ref{res}). With
two DLAs each at $z_{\rm abs}<1$ and $z_{\rm abs}>2$, we prioritised
according to highest neutral hydrogen column density, the largest
background radio flux ($>1$ Jy) and the most compact background source
size, in order to maximise the covering factor. Unfortunately, this
sample, prioritised by the first two criteria, gave a range of values
for the radio source size ($\theta_{\rm QSO}\approx1'' - 15''$, see
Table \ref{size}), none of which are especially compact. However, in
order to shed light on the above arguments, DLAs occulting all radio
source sizes should be searched. Furthermore, at the time of
application, the DLA at $z_{\rm abs}=0.6561$ towards 1622+238 was
believed to be due to a spiral galaxy (\citealt{clwb98}, although see
\citealt{ctm+07}) and the background radio source size of $15.3''$ is
similar to that of the other spirals detected in 21-cm absorption
(\citealt{cmp+03} and references therein).

\subsection{Observations, Data Reduction and Results}

All of the observations were performed in June and July of 2006 with
the Westerbork Synthesis Radio Telescope in the Netherlands. To cover
the redshifted 21-cm line, the UHF and 92-cm receivers were backed by
a band-width of 5 MHz over 2048 channels (dual polarisation), giving
channel spacings of 0.85 to 1.9 \kms ~(for $z_{\rm abs} = 0.66$ to
2.7). This ensured that the observations not only covered
uncertainties in the optical redshifts, but gave a fine enough
resolution to avoid resolving out any possibly narrow absorption
lines; the full-width half maxima (FWHMs) range from 4 to 53
\kms ~for the DLAs already detected in 21-cm (see
\citealt{cmp+03}). The two orthogonal polarisations (XX \& YY) were
used in order to allow the removal of any polarisation dependent radio
frequency interference (RFI). Upon the removal of this, the
polarisations were combined in order to maximise the signal-to-noise
ratio. For all of the observations, the quasars 3C\,48, 3C\,147 and
3C\,286 were used for bandpass and flux calibration. The data were
reduced using the {\sc miriad} interferometry reduction package, with
which we extracted a summed spectrum from the emission region of the
continuum maps.

\subsubsection{0149+336} 
\begin{figure}
\vspace{6.2cm}
\includegraphics{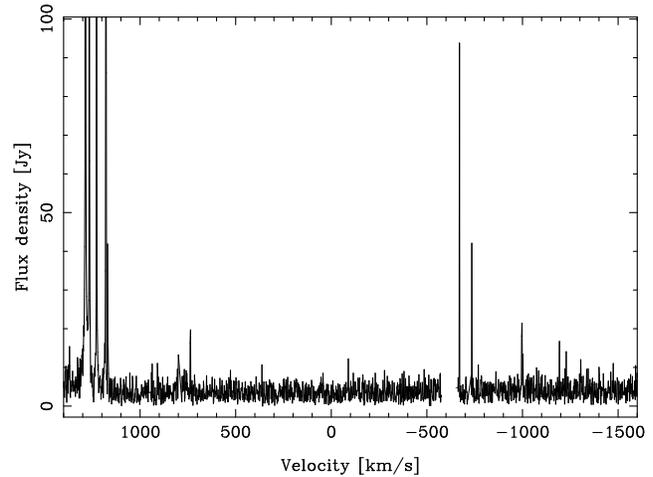}
%line join xfigged out
\caption{The baseline average of the YY polarisation for 0149+336. The
velocity offset is relative to 452.17 MHz ($z=2.14131$) and ranges
from 450.0 to 454.6 MHz ($z=2.1245-2.1565$).}
\label{0149}
\end{figure}

0149+336 was observed in $29\times0.42$ hour slots on 25 June 2006,
with the UHF-low receiver.  However, in this band the RFI was severe,
particularly in the XX polarisation. Upon flagging this and
the worst affected channels out of the YY data, RFI still dominated
on each baseline at all time intervals and the remaining data were of
too poor a quality to obtain an image (Fig. \ref{0149}).

\subsubsection{0809+483} 
 
\begin{figure}
\vspace{6.2cm}
\includegraphics{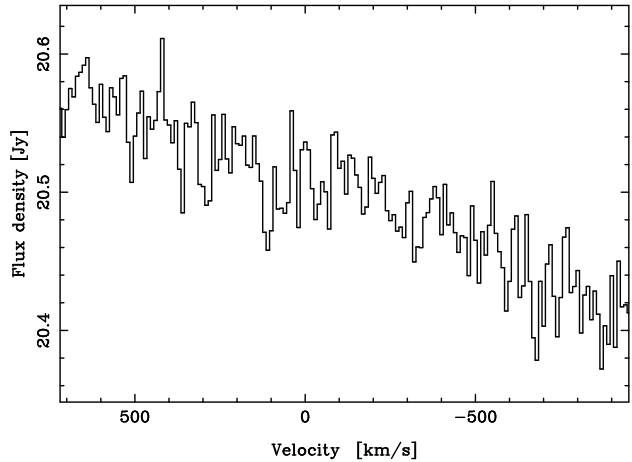}
\caption{Spectrum extracted from the cube of 0809+483. The velocity
offset is relative to 785.68 MHz ($z=0.80787$) and ranges from 783.8
to 788.2 MHz ($z=0.8021-0.8122$). The r.m.s. noise
is 110 mJy per each 0.93 \kms\ channel. In this and Figs. \ref{1402} and
\ref{1622} the data have been redressed to a resolution of 10 \kms.}
\label{0809}
\end{figure}

After realising that 0149+336 was mistakingly observed in 10 second
integrations, we switched to the default 60 seconds, making each slot
2.5 hours long, for which 0809+483 and the other remaining sources
were observed for 5 slots. 0809+483 (3C\,196) was observed on 2 July
2006, with the UHF-high receiver.  Severe RFI in this band (783.8 to
788.2 MHz) meant that one of the slots had to be removed
completely. Furthermore, for all of the slots, the XX polarisation had
to be completely flagged from the baselines involving antenna 3,
leaving 76 full and partial (XX or YY) baseline pairs, over a total
observing time of 7.5 hours, after the flagging of further time
dependent RFI. The source was unresolved by the $66''\times39''$
beam and the final extracted spectrum is shown in Fig.~\ref{0809}.

\subsubsection{1402+044} 
\begin{figure}
\vspace{6.2cm}
\includegraphics{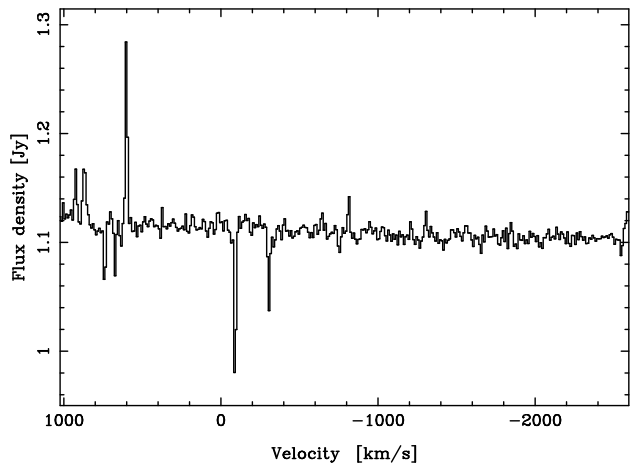}
\caption{Spectrum extracted from the cube of 1402+044. The velocity
offset is relative to 383.06 MHz ($z=2.70805$) and ranges from 381.7
to 386.3 MHz ($z = 2.677-2.721$). The r.m.s. noise
is 32 mJy per each 1.9 \kms\ channel.}
\label{1402}
\end{figure}

1402+044 was observed on 1 July 2006 with the 92-cm receiver. Although
the strongest absorber towards this quasar would exhibit 21-cm
absorption at $\approx383.06$ MHz, the central frequency was offset
from this (as shown in Fig. \ref{1402}) in order to also cover the other two
absorbers in this band (see Table \ref{res}).  The observations were
RFI free, except the last 2.5 hour slot, which was rejected along with
the calibrator used after the run, 3C\,48.  This left 9.3 hours of
good data and the RFI-free observations of 3C\,286 were used for the
bandpass and flux calibration. Thanks to the low degree of
interference in this band, 91 full baseline pairs could be used.  The
source was unresolved by the $708''\times72''$ beam.

All but one of the features in the extracted spectrum
(Fig. \ref{1402}) were noted to be due to RFI during the reduction
process and occurred in one polarisation only. The exception is the
``absorption feature'' closest to $v=0$ \kms\ (Fig. \ref{1402}) and
thus the only confirmed DLA (with $N_{\rm HI}=8\times10^{20}$ \scm,
Table~\ref{res}). This feature was present in both polarisations, each
of which can be fitted by similar Gaussians, giving $\Delta v = -90.2$
\& $-89.3$ \kms\ and FWHMs of $12.7$ \& $11.2$ \kms\ for the XX and YY
polarisations, respectively. These values agree to less than one
channel width and give a redshift of 2.707, cf. the quoted DLA value
of $z_{\rm abs}= 2.708$ \citep{ph04}, $z_{\rm abs}= 2.7069\pm0.0003$
(Si{\sc \,ii}, 1526 \AA) and $z_{\rm abs}= 2.7072\pm0.0003$ (Al{\sc
\,ii}, 1670 \AA), from the Sloan Digital Sky Survey. However, the fact
that the depth of the feature differs significantly between each
polarisation ($-0.26$ Jy in XX \& $-0.13$ Jy in YY), as well as these
features also appearing at other locations in the image remote from
the continuum source, forces us to conclude that this absorption
feature is an artifact.  
%%%%{\bf but 1402+044 is a BL Lac - check out
%%%%this source is A\&A 325, 109: ``Are high polarization quasars and BL
%%%%Lacertae objects really different? A study of the optical spectral
%%%%properties.'' Surely polarisation would be measured in the flux, not
%%%%the absorption?\\ YLVA IS RE-ANALYSING THIS. ASSUME NON-DETECTION FOR
%%%%THE MOMENT}

\subsubsection{1622+238}
\begin{figure}
\vspace{6.2cm}
\includegraphics{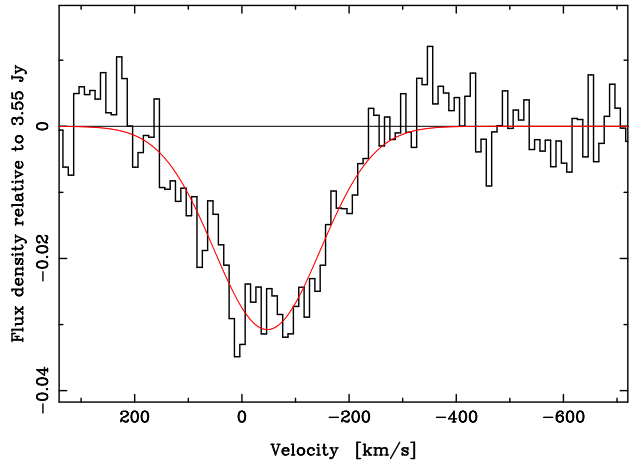}
\caption{Spectrum extracted from the cube of 1622+238. The velocity
offset is relative to 857.681 MHz ($z=0.6561$) and ranges from 856.6
to 860.1 MHz ($z = 0.6514-0.6582$). The r.m.s. noise
is 17 mJy per each 0.85 \kms\ channel. The Gaussian fit gives a peak
line depth of $31$ mJy, a centroid at -47 \kms\ and a FWHM of 235
\kms\ for the line.}
\label{1622}
\end{figure}

1622+238 (3C\,336) was observed on 9--10 July 2006 with the UHF-high
receiver. After flagging of time dependent RFI, 12.0 hours of good
data remained, although there was some RFI remaining on some
baselines, particularly in the XX polarisation. After further
flagging, 63 full and partial pairs remained, resulting in a
detection (Fig. \ref{1622}).
Further details are given in \citet{ctm+07}, where this detection
is reported and discussed.

\section{Discussion}

%\subsection{Comparison with previous results}

\subsection{These and other recent results}
\label{these}

In Table \ref{res} we present our observations and derived results.
\begin{table*}
\centering
%\begin{minipage}{175mm}
\begin{minipage}{170mm} %times
\caption{Our search results. $z_{\rm abs}$ is the redshift of the DLA
with the optical identification (ID) given: S--spiral \citep{clwb98},
U--unknown. $z_{\rm em}$ is the redshift of the background QSO, $S$ is
the flux density at the observed frequency, $\nu_{\rm obs}$, $S_{\rm
HI}$ is the peak flux of the line and $\Delta v$ is the channel width
of the observations. Since the measured r.m.s. noise is dependent upon
the spectral resolution, as per \citet{cmp+03}, for the non-detections
the $3\sigma$ upper limits of $\tau_{\rm peak}$ at a velocity
resolution of 3 \kms\ are quoted, since this is a fairly typical
resolution for the previous and new \citep{gsp+06} searches. The total
neutral hydrogen column density, $N_{\rm HI}$ [\scm], is given with
the corresponding reference, which yields the quoted spin
temperature/covering factor ratio. In the case of the non-detections,
like \citet{cmp+03}, we assume a FWHM of 20 \kms\ for the line width (the mean value of
the detections, excluding 1622+238). \label{res}}
\begin{tabular}{@{}l c c c c c c c  c r c r@{}} 
\hline
QSO & $z_{\rm abs}$ & ID & $z_{\rm em}$ &$\nu_{\rm obs}$ [MHz] & $S$ [Jy] & $S_{\rm HI}$ [mJy] & $\Delta v$  [\kms] &  $\tau_{\rm peak}$ &$\log_{10}N_{\rm HI}$ & Ref. & $T_{\rm spin}/f$ [K]\\
\hline
0149+336&  2.1413 & U & 2.431 & 452.17   & -- & --&1.6 &-- & 20.6 & 1,2 &--\\
0809+483$^a$ & 0.80787 &  U  & 0.871 &785.68  & 20.5 & $<110$& 0.93& $<0.0090$& $>20.3$ & 3 & $>600$\\
1402+044  & 2.688$^b$ & U & 3.215&  385.14 & 1.12 &  $<32$  & 1.9 &$<0.068$ & -- & 2 & --\\
... & 2.708 & U  & ...& 383.07 & ...&...& ...& ...& 20.9 & 4 &$>300$\\
... & 2.713$^b$ & U & ...& 382.55  & ...&...& ...& ...& -- & 2 & --\\
1622+238$^c$ & 0.6561 & S &  0.927 & 857.68 & 3.55&   $31$ &0 .85  &0.0088  & 20.4 & 5 & 60\\
\hline
\end{tabular}
{Notes: $^a$3C\,196 (candidate DLA at $z_{\rm abs}=0.808$),
$^b$candidate sub-DLAs \citep{twl+89}, although possibly due to blends of narrower
features (figure 3 of \citealt{wtsc86}),
$^c$3C\,336. \\ References: (1) \citet{wlfc95}, (2) \citet{twl+89},
(3) \citet{lvm98}, (4) \citet{ph04}, (5) \citet{rt00}.}
\end{minipage}
\end{table*}
 In the optically thin regime ($\sigma/f.S\lapp0.3$),
 Equation~\ref{enew} reduces to $N_{\rm
 HI}=1.823\times10^{18}\frac{T_{\rm spin}}{f}\int\!\frac
 {\sigma}{S}\,dv\,,$ thus giving a direct measure of the spin
 temperature of the gas for a known column density and covering
 factor.  However, as recalled in the introduction, in the absence
 of any direct measurement of the size of the radio absorbing region,
 this latter value cannot be determined.  Therefore in Table
 \ref{res}, we quote our derived results in terms of $T_{\rm spin}/f$.

\begin{table*}
\centering
%\begin{minipage}{135mm}
\begin{minipage}{130mm}%times
\caption{Summary of the new (since \citealt{cmp+03}) 21-cm absorption
searches in DLAs ($N_{\rm HI}\ge2\times10^{20}$ \scm) and sub-DLAs
($N_{\rm HI}<2\times10^{20}$ \scm). In the top panel we list the
detections and in the bottom panel the non-detections (quoted as
$3\sigma$).}
\label{new}
\begin{tabular}{@{}l c c  c c c @{}} 
\hline
Reference      & No.& $z_{\rm abs}$ &  $z_{\rm em}$ & $N_{\rm HI}$ [\scm] & ${T_{\rm spin}}/{f}$ [K]\\
\hline
Previous detections$^{a}$ & 15 & 0.09--2.04 & 0.64--2.85 & $0.2 - 6\times10^{21}$ & $100 - 10,000$\\ 
\citet{kse+06} & 1 & 2.347 & 2.852 & $6\times10^{20}$ & 1500\\
\citet{gsp+06}$^{b}$ & 3 & 1.17--1.37 & 1.37--1.64 & $0.4 - 2\times10^{18}.({T_{\rm spin}}/{f})$ & unknown\\
\citet{kcl06}$^{c}$ & 1 & 3.386 & 3.610 & $2.5\times10^{21}$ & 1600\\
%ZWANN RESULTS? & & & & \\
This paper  & 1 & 0.6561& 0.927& $2.3\times10^{20}$ & 60\\
\hline
Previous non-detections$^{a}$ & 16 & 0.10--3.18 & 0.31--3.61  & $0.1 - 3\times10^{21}$ & $>200\, - > 6000$\\
\citet{gsp+06}$^{b}$ & 6 & 1.31--1.45 & 1.40--2.17 & $<2 - 4\times10^{17}.({T_{\rm spin}}/{f})$ & unknown\\
\citet{sgp06} & 1 & 1.365 & 2.22 & $2\times10^{19}$  & $>900$\\
%ZWANN RESULTS? & & & & \\
This paper  & 2 & 0.81--2.71& 0.87--3.22& $>2 - 8\times10^{20}$& $>300\, - >600$\\
\hline
\end{tabular}
{Notes: $^{a}$See \citet{kc02,cmp+03} for full details. Note that
0438--436 has now graduated from a non to a detection
\citep{kse+06}. $^{b}$DLA candidates only and so neutral hydrogen
column densities are unavailable, hence the upper limits in $N_{\rm HI}$
for the non-detections. Note that the sub-DLA 0237--223 is
common to both \citet{gsp+06} and \citet{sgp06}. $^{c}$We quote the most recent
result as this was previously detected by \citet{dob96,bbw97}.}
\end{minipage}
\end{table*}

\begin{figure}
\vspace{6.5cm}
\includegraphics{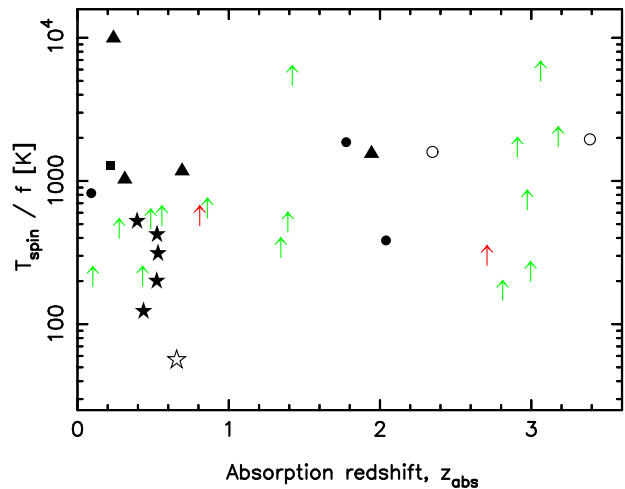}
\caption{The spin temperature/covering factor ratio versus the
absorption redshift for the DLAs searched in 21-cm absorption. The
symbols represent the 21-cm detections and the shapes represent the
type of galaxy with which the DLA is associated: circle--unknown type,
star--spiral, square--dwarf, triangle--LSB. The arrows show the lower
limits and all of these bar one (0454+039 at $z_{\rm abs}=0.8596$)
have unknown host identifications (our values are shown in red, Table
\ref{res}). The unfilled symbols show the new detections since
\citet{cmp+03}, from which the figure is updated: The two new high
redshift detections of \citet{kse+06,kcl06} and the possible spiral
towards 1622+238 \citep{ctm+07}.}
\label{Toverf}
\end{figure}
In Fig. \ref{Toverf} we add these and the other recent results to the
$T_{\rm spin}/{f}$--redshift distribution of \citet{cmp+03}. Like
\citet{kc02}, for the detections we see that at low redshift we have
a mix of ratios (cf. spin temperatures) and at high redshift
exclusively high values, although not segregated to the extent seen in
figure 3 of \citet{kc02}.  We also see that the low values are
dominated by the identified spirals, although by no means does this
irrefutably indicate that the lower redshifts do indeed have a mix of
spin temperatures: If the larger spirals have larger covering factors
than the dwarfs and LSBs, then the spin temperatures of
these non-spirals would be correspondingly lower. 

As stated, however, in view of the uncertainty in determining the
covering factor and spin temperature, we prefer to retain the
degeneracy. Through the two new high redshift end points
\citep{kse+06,kcl06}, which nearly double the number of detections at
$z_{\rm abs}>1$, it appears as though $T_{\rm spin}/{f}$ does not
increase indefinitely with  redshift.  This supports our
previous hypothesis that 21-cm absorption should be readily detectable
towards compact radio sources at $z_{\rm abs}\gapp2$
\citep{cmp+03}. Since $f\leq1$, the ordinate values in
Fig. \ref{Toverf} can be considered as the maximum permissible values
of $T_{\rm spin}$, and so the very high spin temperatures at $z_{\rm
abs}\gapp3$ \citep{kc02} are not seen.  Ironically, the largest
$T_{\rm spin}/{f}$ value (10,000 K)\footnote{Detected by \citet{kc01a}
and using the total neutral hydrogen column density of $N_{\rm
HI}=2\times10^{21}$ \scm, obtained from Hubble Space Telescope
observations \citep{rt00}.} is obtained in one of the lowest redshift
DLAs of the sample, at $z_{\rm abs}=0.238$ towards 0952+179. 
%Apart from this example, Fig.~\ref{Toverf} suggests that spin temperatures do not generally get far above $T_{\rm spin}\approx2000$ K.

%\footnote{Note that, upon its detection, the previous lower limit of $T_{\rm spin}>3300$ K for 0210+113 \citep{kc02} has now decreased to an upper limit of $T_{\rm spin}\leq1950$ K (for $f\leq1$).}.

\subsection{The detection of 21-cm absorption}
\subsubsection{Radio source sizes}

If $T_{\rm spin}/f$ generally levels off at high redshift, as
suggested by Fig.~\ref{Toverf}, we reiterate that 21-cm absorption
should be readily detectable (especially towards compact radio
sources, thus maximising $f$). In Table~\ref{size} we show the sizes
of the background continuum sources, as obtained from the highest
resolution radio images available, closest in frequency to the
redshifted 21-cm values.
\begin{table}
%\centering
%\begin{minipage}{87mm}
\begin{minipage}{80mm}%times
\caption{The radio source sizes, $\theta_{\rm QSO}$ [arc-sec] at
$\nu_{\theta}$ [GHz], of the QSOs illuminating the recent DLAs searched (Table
\ref{new}). \label{size}}
\begin{tabular}{@{}l c  c  c  l @{}} 
\hline
QSO & Study & $\theta_{\rm QSO}$ & $\nu_{\theta}$ & Reference \\
\hline
J0108--0037 & G07 & 0.98&1.4 & FIRST\\
B0201+113 & K07 & 0.018 & 0.33 & K07\\
B0438--436 & K06 & 0.039 & 1 \& 5 & VLBI/VLA$^{\dagger}$\\
J0804+3012 & G07 &$\approx22,<16.5$& 0.65,1.4 & G07,NVSS\\
B1622+238 & C07 & 15.33 & 1.4 & FIRST\\
J2358--1020 & G07 &0.95& 1.4& FIRST \\
\hline
%B0149+336 & C07  &15.4 & 1.4 & NVSS\\
J0214+1405 & G07 &$<18.8$&1.4 & NVSS\\
J0240--2309 & G07 &$<18.8,<0.05$&1.4,2.3 & NVSS,EVN\\
J0748+3006& G07 &1.93& 1.4& FIRST\\
B0809+483 & C07 &5.27  &  0.4 & VLA/Merlin$^{\dagger}$\\
J0845+4257& G07 &1.06& 1.4& FIRST\\
J1017+5356 & G07 & 3.84& 1.4& FIRST\\
B1402+044  & C07 & 1.35  & 1.4 & FIRST \\
J1411--0300& G07 &$\approx25,6.98$&0.59,1.4 & G07,FIRST\\
J1604--0019 & G07 &$\approx37,2.13$& 0.61,1.4& G07,FIRST\\
%...& 23.6 & 1.4 & NVSS\\
\hline
\end{tabular}
{References: K06: \citet{kse+06}, G07: \citet{gsp+06}, K07:
\citet{kcl06}, C07: This paper. VLBI: Very Long Baseline
Interferometry ($^{\dagger}$see table 2 of \citealt{cmp+03}) for
details. EVN: (European VLBI Network \citep{dbam98}, FIRST: The Very
Large Array's ``Faint Images of the Radio Sky at Twenty
Centimetres'', NVSS: ``NRAO VLA Sky Survey''}.
\end{minipage}
\end{table}
From this, we see that the two new high redshift end points (0201+113
\& 0438--436) occur towards the most compact background radio sources
(which are also compact in comparison to all of the DLAs searched,
table 2 of \citealt{cmp+03}). This may indicate that the spin
temperatures in these two DLAs, whose host types are unidentified, are
indeed higher than in the low redshift spirals, although still
$\lapp2000$~K. However, in the absence of any knowledge of the
absorption cross section, this is still inconclusive.

Apart from these two new high redshift examples, the radio source
sizes do not reveal much, with a very similar range of values between
the detections and non-detections. While the only two significantly
smaller radio sources illuminate DLAs detected in 21-cm, so does the
largest, although the absorber towards 1622+238 may have an impact
parameter $\gapp75$ kpc \citep{sdm+97,clwb98}. The FWHM of 235 \kms\
for the 21-cm absorption, which is over four times wider than any
other DLA and an order of magnitude wider than the average value, also
suggests that this is a very extended absorption system
\citep{ctm+07}.  This emphasises the need for knowledge of the size of
the absorbing region in relation to the background emission, as the
size of the background source by itself gives an incomplete picture of
the covering factor.

\subsubsection{Angular diameter distance ratios}
\label{addr}

As discussed in \citet{cw06}, a major factor in determining the extent
of the absorbing cross section in relation to the background emission
is how much closer the absorber is to us than the emitter. At low
redshifts ($z_{\rm abs}\lapp0.5$, see figure 3 of \citealt{cw06}), the
absorber/quasar angular diameter distance ratio ($DA_{\rm DLA}/DA_{\rm
QSO}$) follows the (near) linear increase with which we are familiar
-- move something twice as far and it appears to be half the
size. However, beyond $z_{\rm abs}\approx1.6$, the angular diameter
distance begins to decrease and so the absorber and quasar are
essentially at the same angular diameter distance for $z_{\rm
abs}\gapp1$. Therefore the DLA gains no advantage in its coverage by
being at a lower redshift than the quasar and in fact, the larger the
redshift of the quasar, the more the coverage is decreased.

As seen from Fig. \ref{Toverf}, most of the non-detections may simply
not have been searched deeply enough\footnote{\label{limit}Since the
ordinate of Fig. \ref{Toverf} is proportional to $(N_{\rm
HI}\,.\,S)/({\rm FWHM}\,.\,\sigma)$, this gives an accurate indication
of the limits, although if the FWHM of the any non-detected 21-cm
absorption is $<20$ \kms (Sect. \ref{these}), the limits would be
better than shown. The plot nevertheless gives a truer representation
of the limits reached than the spin temperature plots of
\citet{ck00,kc01a,kc02}, where a covering factor is assigned.},
particularly if the host galaxies are non-spirals. Statistically,
however, combining the new with the previous results (Table
\ref{new}), at $z_{\rm abs}<1.6$ the binomial probability of obtaining
11 or more detections out of 13 DLAs at $DA_{\rm DLA}/DA_{\rm
QSO}<0.8$\footnote{$DA_{\rm DLA}/DA_{\rm QSO}=0.8$ splits the low
redshift sample approximately in half and is where the increase in
$DA_{\rm DLA}/DA_{\rm QSO}$ with redshift begins to lose its
linearity.}, with 16 or more non-detections out of 22 DLAs, sub-DLAs
and candidates\footnote{The possible candidates at $z_{\rm abs}=2.688$
and $z_{\rm abs}=2.713$ absorbers towards 1402+044 are not included.},
is 0.03\% (cf. 0.06\% previously), a significance of $3.6\sigma$
assuming Gaussian statistics. This becomes 0.02\% when no redshift
partition is used ($\geq11$ out of $13$ detections at $DA_{\rm
DLA}/DA_{\rm QSO}<0.8$ and $\geq24$ out of $35$ non-detections at
$DA_{\rm DLA}/DA_{\rm QSO}>0.8$).

\subsection{The strength of the 21-cm absorption}

\subsubsection{Correlation with Mg{\sc \,ii} equivalent width}
\begin{figure}
\vspace{15.2cm}
%\vspace{14.9cm}
%\special{psfile=3-vel-error.ps hoffset=-15 voffset=450 hscale=73 vscale=73 angle=-90}
\includegraphics{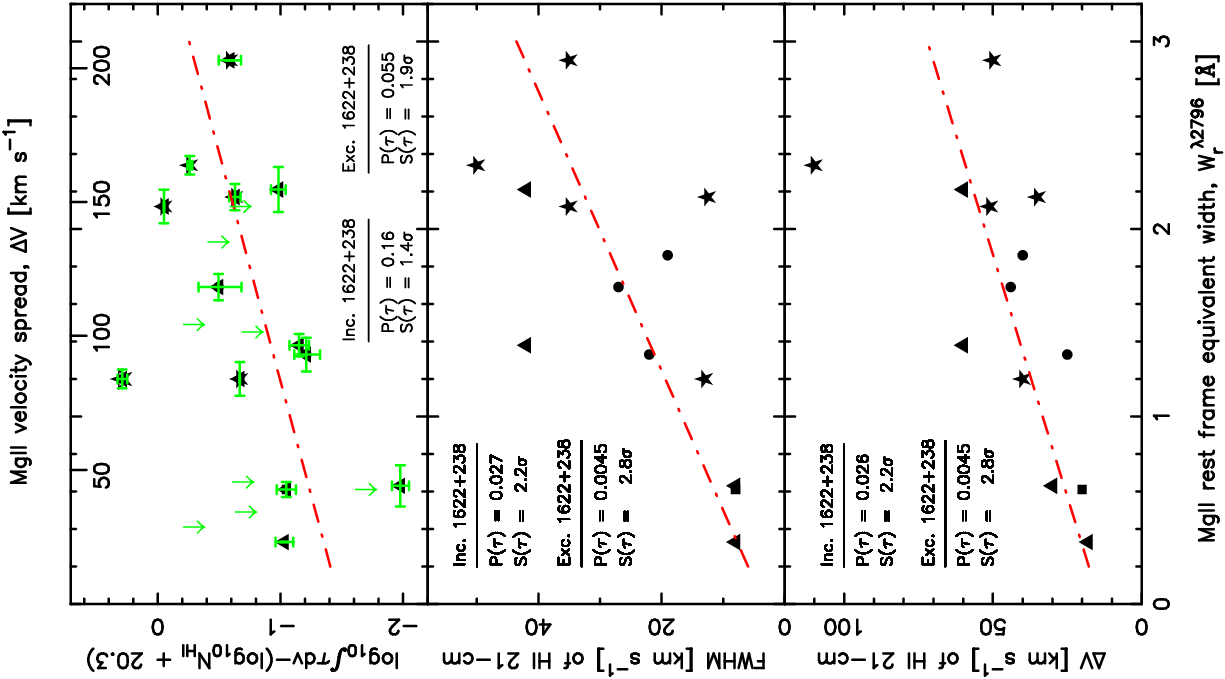}
\caption{The 21-cm line strength and widths versus the rest frame
equivalent width of the Mg{\sc \,ii} 2796 \AA\ line for the DLAs
searched in 21-absorption. As per \citet{mcw+07}, the Mg{\sc \,ii}
velocity spread, shown by the extra markers in the top panel, is given
by $\approx70$ [\kms\ \AA$^{-1}]\times$W$_{\rm r}^{\lambda2796}$
[\AA]. Top: The normalised velocity integrated optical depth (see
\citealt{cmp+03}). The errors on the ordinate are from both the
uncertainties in the column densities and the velocity integrated
optical depths, from the articles cited in \citet{cmp+03}. The errors
on the abscissa are from the articles cited in \citet{mcw+07}. Where
errors in the velocity widths are given, these are usually $\approx1$
\kms. Middle: The FWHM of the 21-cm profile. Bottom: The total
velocity spread of the 21-cm absorption components. In the top panel
the 21-cm non-detections are represented by coloured downward arrows
showing upper limits. In each panel Kendall's $\tau$ two-sided
probability that no correlation exists, $P(\tau)$, is shown along with
the significance of the correlation (derived from this assuming
Gaussian statistics), $S(\tau)$, and the least-squares fit to the
21-cm detections. For clarity, the outlier 1622+238 is only shown in
the top panel (see main text). There are no equivalent width
measurements for 0201+113 nor 0438--436.}
% http://davidmlane.com/hyperstat/z_table.html  ON MART'S COMPUTER FOR SIGNIFICANCE. USE THE BOTTOM APPLET, JUST PUT IN TAU PROB IN SHADED AREA AND SELECT BOTTOM (OUTSIDE) BUTTON 
\label{width}
\end{figure}
Now that we have recapped the factors which could determine whether
21-cm absorption is detected or not, we turn our attention to what
could possibly determine the strength of the absorption, where
detected. \citet{cw06} noted that, although the strength of the
absorption appears to be correlated with the equivalent width of the
Mg{\sc \,ii} 2796 \AA\ line, this is not decisive in whether
absorption is detected or not: In Fig.~\ref{width} (top panel) we see
that the non-detections span a similar range of equivalent widths as
the detections and that 21-cm absorption is detected down to W$_{\rm
r}^{\lambda2796}=0.33$ \AA.  Although there is a strong tendency for
21-cm absorption to occur in DLAs originally identified through the
Mg{\sc \,ii} doublet, \citet{cw06} argue that this is a mere
consequence of the Mg{\sc \,ii} selection bias towards low redshift
absorbers and thus lower angular diameter distance ratios (see their
figure 3). Therefore the deciding criteria in regards to detectability
may be related to the coverage, as discussed above. Of course, if the
fit in Fig. \ref{width} (top) is an accurate indicator, again, the
non-detections may not have been searched sufficiently
deeply$^{\ref{limit}}$ to overcome these low covering factors.

For strong (W$_{\rm r}^{\lambda2796}\gapp0.3$ \AA) Mg{\sc \,ii}
absorption, characteristic of the DLAs searched in 21-cm, the line is
completely saturated and above these equivalent widths the profile
traces the number of absorbing components
(e.g. \citealt{ell06}). Therefore the strength of the Mg{\sc \,ii}
absorption is dominated by the velocity structure. On the other hand,
the 21-cm absorption is optically thin, and therefore less susceptible
to the same kinematics, although the top panel of Fig.~\ref{width}
suggests that this may nevertheless by important in determining the
strength of the 21-cm profile. In order to verify this, in the lower
two panels of Fig. \ref{width} we show the full width half maximum and
the total velocity spread ($\Delta V$) of the 21-cm absorption
profiles (obtained from the references given in table 1 of
\citealt{cmp+03}). The fact that the non-parametric correlations are
considerably more significant than for the line-strength correlation,
suggests that the strength of the 21-cm absorption may indeed be
dominated by the velocity structure\footnote{A far weaker, non-linear
correlation between $\Delta V$ and W$_{\rm r}^{\lambda2796}$, in a
sample of five Mg{\sc \,ii}/21-cm absorbers, was previously noted by
\citet{lan00} [figure 3.5], who interpreted this as the spread of the
21-cm components being related to the Mg{\sc \,ii} equivalent width
and thus the number of Mg{\sc \,ii} velocity components.}, although
the spirals do introduce some scatter, perhaps due to a contribution
from the large-scale galactic dynamics. In particular, 1622+238 (FWHM
= 235 \kms\ and $\Delta V\approx560$ \kms), which is more reminiscent
of a large-scale emission profile, rather than the typical pencil beam
absorption profile \citep{ctm+07}.

\subsubsection{Correlation with metallicity and implications}

\begin{figure}
\vspace{6.5cm}
\includegraphics{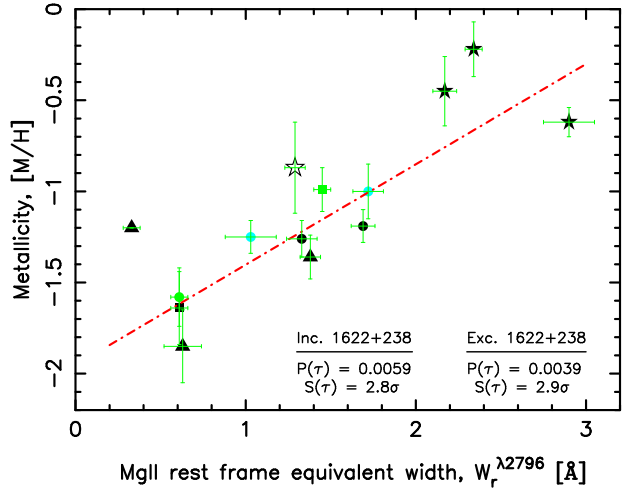}
\caption{The metallicity versus the rest frame equivalent width of the
Mg{\sc \,ii} 2796 \AA ~line for the DLAs illuminated at radio
frequencies, where available. The black symbols represent the 21-cm
detections and the coloured (green/blue) symbols the
non-detections/unsearched DLAs. The errors are from the literature as
given in \citet{mcw+07} and are not shown in the other plots for the
sake of clarity. Since we use the same [M/H] and W$_{\rm
r}^{\lambda2796}$ range in each plot, these give a clear indication of
what the uncertainties are. Even in this sub-sample, we see the
correlation exhibited for all of the 49 DLAs and sub-DLAs (over
$z_{\rm abs}=0.2 - 2.6$) for which both measurements exist
\citep{mcw+07}; in fact it was the above correlation which prompted
this investigation of the general DLA population.  As per
Fig. \ref{width}, in this and the following plots we give the
non-parametric correlation estimator and the resulting significance.}
\label{M-W-radio}
\end{figure}

In addition to the correlation between the Mg{\sc \,ii} and 21-cm
profile widths, a relationship between the velocity spreads of low
ionisation lines and the metallicity has been well documented:
\citet{wp98} originally noted a tentative correlation in a sample of
17 DLAs, over the redshift range $z_{\rm abs}=1.6 - 3.0$, with a
similarly tentative correlation between [Zn/H] and $\Delta V_{\rm
ion}$ being found in a sample of 72, over the range $z_{\rm abs}=1.4 -
4.5$, by \citet{pdd+03}. From composites of 370 SDSS spectra,
\citet{nrtv03} found higher metallicities in the W$_{\rm
r}^{\lambda2796}\geq1.3$ \AA\ sample than for $1.0\leq$W$_{\rm
r}^{\lambda2796}<1.3$, over both low and high redshift
regimes.
This was confirmed by \citet{trn+05} with composites
from nearly 6000 SDSS spectra and metallicities obtained from Zn, Si,
Cr, Fe and Mn abundances. A correlation between the metallicity and
velocity spread from individual systems was found by \citet{lpf+06},
using several low ionisation species (O{\sc \,i}, Si{\sc \,ii}, Fe{\sc
\,ii}, Cr{\sc \,ii} \& S{\sc \,ii}), in 70 DLAs and sub-DLAs (with
$N_{\rm HI}\ge10^{20}$ \scm)\footnote{A correlation between the
metallicity and velocity spread is also found for ``true sub-DLAs''
($10^{19}\ge N_{\rm HI}\ge2\times10^{20}$ \scm) by \citet{mlk+07}.}
over the redshift range $z_{\rm abs}=1.7 - 4.3$. Like \citet{wp98}
[and \citealt{kkp+07,pcw+07}], the velocity spread is attributed to
the galactic dynamics, indicating that the metallicity traces the mass
of the galaxy with which the DLA is associated. However,
\citet{bmp+06} find, from a sample of 1806 Mg{\sc \,ii} absorbers with
W$_{\rm r}^{\lambda2796}\geq0.3$ \AA, that the halo mass and
equivalent width are {\em anti-correlated}, thus leading
\citet{mcw+07} [see Fig. \ref{M-W-radio}] to suggest that the velocity
spread may be due to outflows of cold dusty gas, which enrich the gas
with metals\footnote{Note also that the abundance of metals appears to
be anti-correlated with the neutral hydrogen column density, which
when combined with a mass--metallicity relationship, suggests that
sub-DLAs arise in more massive galaxies than DLAs, perhaps due a
deficit of neutral gas in the central regions of these larger
galaxies.  This may suggest that it is the sub-DLAs which contribute
star forming activity, and thus elemental abundances, to the early
Universe \citep{wp98,kkp+07}, perhaps bypassing the inconsistency
raised by the low abundance of cold, star forming, molecular gas observed
in DLAs (Sect. 1).}. Finally, note that although the gas responsible
for C{\sc \,iv} absorption is not believed to be the same as that
evident through singly ionised absorption \citep{wp00}, a correlation
between the metallicity and C{\sc \,iv} velocity spread has also been
found \citep{flps07}\footnote{Interestingly, in 40\% of the absorbers
the C{\sc \,iv} gas exceeds the escape velocity, suggesting that the
absorption occurs in outflows. Unlike, the outflows of \citet{mcw+07},
however, these consist of photoionised and collisionally excited gas.}.

Metallicity--velocity spread correlations are therefore common to both
DLAs and sub-DLAs in a variety ionisation states. The fact the 21-cm
line strength and width are both correlated with the velocity spread
of the Mg{\sc \,ii} therefore suggests that these should also follow
the metallicity:
\begin{figure}
 \vspace{9.2cm} \includegraphics{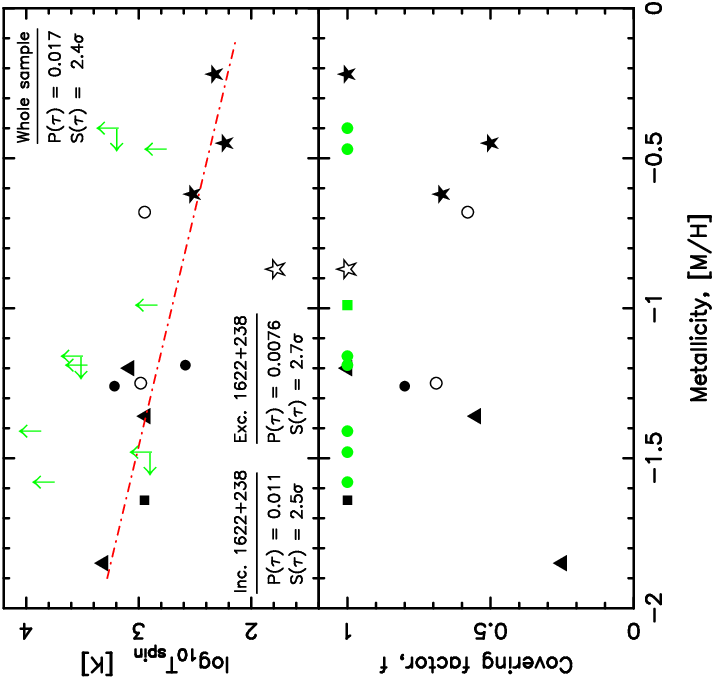} \caption{As figure 1 of
 \citet{cmp+03}, which shows the spin temperature against absorption
 redshift as given in \citet{kc02}, but with metallicity shown on the
 abscissa. In this and Figs. \ref{Toverf-m} and \ref{M-z}, the other
 new measurements are for 0201+113 \citep{epss01} and 0438--436
 \citep{aeps05} and, as per Fig. \ref{M-W-radio}, the metallicities
 are from the references given in \citet{mcw+07}, with the few
 available limits added from \citet{aeps05,ell06}. Kendall's $\tau$
 two-sided probability is shown for both the detections only (with and
 without 1622+238) and the detections + the non-detections (the
 least-squares fit is for the detections only). In the bottom panel,
 which shows the covering factors applied to derive the spin
 temperatures \citep{kc02,kse+06,kcl06}, the coloured symbols
 represent the non-detections (which show that $f$ is set to unity in
 all cases). For 1622+238 we have assumed $f=1$, since
 $f<1\Rightarrow T_{\rm spin}<60$ K in this DLA \citep{ctm+07}.}
\label{kanekar} 
\end{figure} 
\citet{kb04} state that, due to redshift evolution, there is an
anti-correlation between the spin temperature and metallicity in DLAs,
where the higher redshift absorbers have lower metallicities,
resulting in higher spin temperatures \citep{ck00}. Plotting this
(Fig. \ref{kanekar}), we see that with $S(\tau)=2.7\sigma$, the
correlation is fairly significant, although treating the
non-detections as detections decreases this to $2.4\sigma$. Since all
but two of these are located far above the fit, the actual spin
temperatures will be larger, further reducing this
significance\footnote{Although the two non-detections with upper
limits to their metallicities could be located closer to the fit, the
metallicities would have to be ${\rm [M/H]}\lapp-2$ if taking these
limits as the spin temperature values.}. The trend is, however,
expected on the basis that metals provide radiation pathways
(e.g. \citealt{whm+95}), meaning that a higher metallicity gas is
expected to have a higher cooling rate. Furthermore, the metallicity
is known to correlated with the molecular hydrogen abundance, which
traces cold gas, at high redshift \citep{cwmc03,plns06}, further supporting a
spin temperature--metallicity anti-correlation. This suggests that,
for the detections at least, the covering factor estimates may be
reasonable (Fig. \ref{kanekar}, bottom panel), although it leads us to
reiterate that the very high ``spin temperatures'' in the
non-detections could in fact be due to overestimates in the covering
factors of these DLAs.

\begin{figure}
\vspace{6.5cm}
\includegraphics{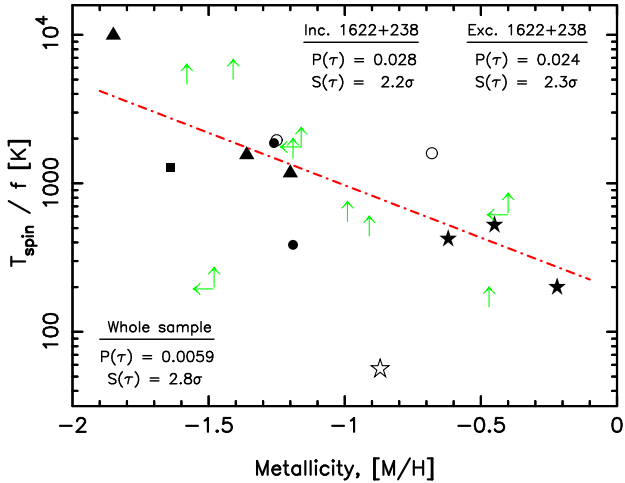}
\caption{The spin temperature/covering factor ratio versus the
metallicity for the DLAs searched in 21-cm absorption. As in
Fig. \ref{kanekar}, Kendall's $\tau$ two-sided probability is shown
for both the detections only (with and without 1622+238) and the
detections + the non-detections.}
\label{Toverf-m}
\end{figure}

As discussed previously, since the covering factor can at best be
 estimated for a very few sources, the spin temperature--covering
 factor degeneracy is best left intact, thus only utilising the values
 of {\em known} parameters. Plotting this (Fig. \ref{Toverf-m}), we
 see that
there is more scatter than in the $T_{\rm spin}$ only plot, resulting
in a slightly lower significance.  However, in this case the
non-detections have considerably less scatter and the fit suggests
that we may be close to the sensitivities required to detect 21-cm in
these DLAs. It should be borne in mind, however, that these are still
lower limits, although they are, on the whole, several factors lower
than the ``spin temperatures'' implied by Fig. \ref{kanekar}, thus
being more consistent with the implied correlation. More to the point,
through retaining the $T_{\rm spin}/f$ degeneracy, only observed
measurements are used.

\begin{figure}
\vspace{6.5cm}
\includegraphics{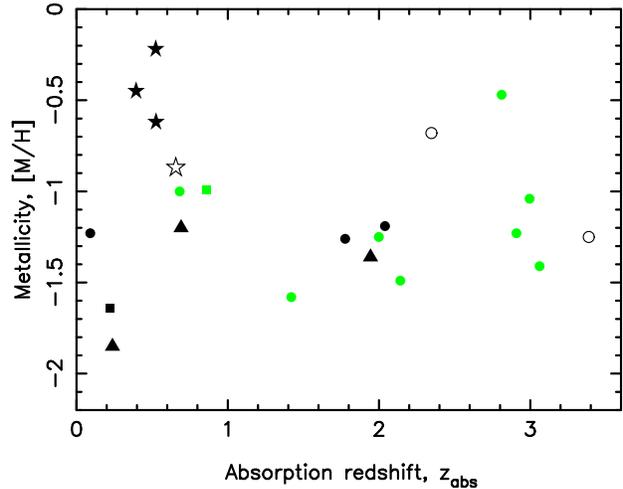}
\caption{The metallicity versus the absorption redshift for the DLAs
under discussion. No overwhelming trend is seen for this sample,
verifying that the relation seen in Fig. \ref{Toverf-m} is not
redshift dependent. Although, like Fig. \ref{Toverf-m}, the spirals
are grouped together at high metallicities.}
\label{M-z}
\end{figure}

From the enrichment of the interstellar medium by successive
generations of stars, a metallicity--redshift anti-correlation is
expected. In conjunction with a possible spin temperature--metallicity
correlation (Fig. \ref{kanekar}), this could explain an increase in
the spin temperature with redshift \citep{kc02}. However, from the
results of \citet{cmp+03,cw06} and the flattening of $T_{\rm
spin}/{f}$ at $\approx2000$ K (Fig. \ref{Toverf}), we believe that
this may be artificial. In Fig.~\ref{M-z} we show the
metallicity--redshift distribution for this sample, which due to the
limited dataset, does not exhibit the metallicity--redshift relation,
present in much larger \citep{pgw+03} or more homogeneous
\citep{cwmc03} datasets. So, although a spin temperature--metallicity
correlation may be expected, we suggest that this is not due to the
spin temperature evolving with redshift.

\subsubsection{The cause of the correlations}

In the optically thin regime, the normalised
21-cm line strength is 
\begin{equation}
\frac{\int\tau\,dv}{N_{\rm HI}}\propto \frac{f}{T_{\rm spin}}, 
\label{equ2}
\end{equation}
where here $\tau\equiv\sigma/S$ (Sect. \ref{these}). Therefore one has
 to be wary of over-interpreting the spin temperature (as well as the covering
 factor): According to Equation \ref{equ2}, Fig. \ref{width} (top) exhibits an
 increase in $f/T_{\rm spin}$ with the Mg{\sc \,ii} rest frame
 equivalent width, which when combined with the
 metallicity--equivalent width relation (Fig.~\ref{M-W-radio}), would
 suggest that $f/T_{\rm spin}$ is correlated with metallicity, giving
 the anti-correlation between $T_{\rm spin}/f$ and [M/H]
 (Fig.~\ref{Toverf-m}).

\begin{figure}
 \vspace{9.2cm} 
\includegraphics{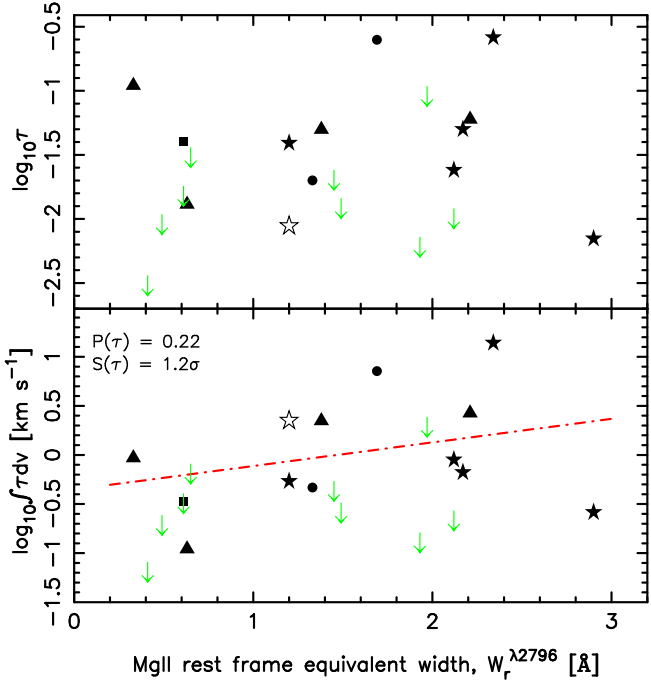} 
\caption{As per Fig. \ref{width} but
 with the 21-cm peak and velocity integrated optical depth on the
 ordinate. The statistics are for the detections only.}  
\label{tau-W}
 \end{figure}

However, $f/T_{\rm spin}$ is but a measure of the normalised 21-cm
 line strength (Equation \ref{equ2}), which, as seen from
 Fig. \ref{width} (middle, bottom), may be a consequence of the
 kinematics, as is the Mg{\sc \,ii} equivalent width. So what we are
 seeing here is that the velocity structure of the cold neutral atomic
 gas does share a degree of coupling with that of the singly ionised
 species\footnote{Such a coupling of the cold atomic and molecular gases
 towards reddened quasars has also recently been found by
 \citet{cdbw07}.}. This confirms that the neutral and singly ionised
 gas are spatially coincident, as suggested by the correspondence of
 their strongest absorption components \citep{tmw+06}.  Therefore, any
 correlation with the metallicity could well be dominated by the left,
 rather than right, hand side of Equation \ref{equ2}, more
 specifically $\int dv$, which, through its tracing of the Mg{\sc
 \,ii} velocity spread will also trace the metallicity.  In order to
 determine which term on the left hand side of the equation is
 dominating the correlation with W$_{\rm r}^{\lambda2796}$, and thus
 [M/H], in Fig.~\ref{tau-W} (top) we show the optical depth of the
 21-cm absorption against the Mg{\sc \,ii} rest frame equivalent width
 and, as noted by \citet{bw83}, there appears to be little correlation
 between these two quantities. It therefore appears that the
 relationship between the 21-cm line strength and the Mg{\sc \,ii}
 equivalent width (Fig. \ref{width}, top), is predominately due to the
 21-cm velocity spread (Fig. \ref{width}, middle \&
 bottom)\footnote{In fact, including the optical depth on its own
 actually has a destabilising effect on the correlation;
 Fig. \ref{tau-W} (bottom) cf. Fig. \ref{width} (middle \& bottom),
 which is somewhat neutralised by normalising this by the column
 density (Fig. \ref{width}, top).  Interestingly, although DLAs at
 $z<1.65$ arise almost entirely from the Mg{\sc \,ii} absorbers with
 ${\rm W}_{\rm r}^{\lambda2796}\geq0.6$ \AA, \citet{rt00} also find
 little direct correlation between $N_{\rm HI}$ and ${\rm W}_{\rm
 r}^{\lambda2796}$.}.

This confirms (as shown by the statistics in Fig. \ref{width}) that
 the kinematics of the 21-cm line is the key factor in giving a
 correlation between the 21-cm line strength and the Mg{\sc \,ii}
 equivalent width, and therefore the metallicity. How this itself ties
 in with the right hand side of Equation \ref{equ2} is more
 complicated, since, as discussed at length above, it is very hard to
 determine the degeneracy between these two unknowns. For example, a
 larger covering factor could well be manifest in larger observed
 velocity spreads of the 21-cm profiles, as seen for the spirals
 (Fig. \ref{width}), particularly 1622+238 \citep{ctm+07}. Conversely,
 the spin temperature could be anti-correlated with the velocity
 spread, although this would be counter-intuitive if the spin
 temperature is related to the kinetic temperature of the gas, a
 problem further compounded if the relationship in Fig.~\ref{kanekar},
 rather than Fig. \ref{Toverf-m}, were used.  On the other hand, as
 expected, the spin temperature could indeed be directly
 anti-correlated with the metallicity (through higher cooling rates),
 or the covering factor correlated, due to larger covering factors
 being associated with larger, more evolved galaxies. Either, or both,
 of these scenarios could explain the relation seen in
 Fig. \ref{Toverf-m}, although we know that for this sample the ratio
 between these two parameters shows no significant redshift evolution
 (Fig. \ref{Toverf}).

In light of these various possibilities, once again it is
prudent to leave the degeneracies intact and only conclude as far as
\begin{equation}
\frac{\int\tau\,dv}{N_{\rm HI}}\propto \frac{f}{T_{\rm spin}} \propto {\rm W}_{\rm r}^{\lambda2796}\propto {\rm [M/H]}, 
\label{equ3}
\end{equation}
where the first and third terms are essentially $\Delta V_{\rm 21-cm}$
 and $\Delta V_{\rm MgII}$, respectively.  
%To conclude, unlike \citet{bw83}, we {\em can} argue that the gas seen in absorption at 21-cm is the same gas which creates the optical absorption lines.

\section{Summary}

We have searched four sight-lines for 21-cm absorption in four DLAs
and two candidate sub-DLAs with the Westerbork Synthesis Radio Telescope. This
has resulted in one detection and four non-detections, with one
observation being lost to severe radio frequency interference at 452
MHz. Adding these to the results from other surveys since
\citet{cmp+03}, gives a total of six new detections (three in
confirmed DLAs) and eleven new non-detections (five in confirmed
DLAs). From these and the previous results, we find:
\begin{enumerate}
  \item There is indeed a mix of spin temperature/covering factor
    ratios at low redshift, although at $z_{\rm abs}\gapp1$ these are
    mostly close to the (typical\footnote{There is one outlier with a
    significantly larger ratio, $T_{\rm spin}/{f}=10,000$~K, but at
    redshift of $z_{\rm abs}=0.238$ this definitely does not
    contribute to an increase in the spin temperature with redshift.})
    maximum value of $T_{\rm spin}/{f}\approx2000$ K. This levelling
    off of the spin temperature at high redshift, confirmed
    by the two new high redshift detections \citep{kse+06,kcl06},
    is consistent with the suggestion by \citet{cmp+03} that the spin
    temperature does not increase indefinitely with redshift.

    \item Occulting extremely compact radio sources ($\leq0.04''$),
      these two new high redshift detections vindicate the prediction of
      \citet{cmp+03} that, despite extremely high ``spin
      temperatures'', DLAs should be detectable at high
      redshift, particularly through the targetting of those located
      towards compact radio sources. However, the radio sources
      illuminating the other new (intermediate redshift) detections,
      are not significantly smaller than those illuminating the
      non-detections ($\sim1''$).
     
\item The addition of the new search results increases the
significance that the 21-detections generally have smaller angular
diameter distances than their background quasars, thus maximising the
covering factor through line-of-sight geometry \citep{cw06}.  This
results in a mix of distance ratios at low redshift, but exclusively
high ratios at high redshift, as per the ``spin temperature''
distribution of \citet{kc02}.

%\end{enumerate}

As suggested by \citet{cw06}, since DLAs originally identified through
 the Mg{\sc \,ii} doublet generally have lower redshifts ($0.2\leq
 z_{\rm abs}\leq2.2$, in the optical bands of ground based telescopes)
 than those identified through the Lyman-\AL\ line ($z_{\rm
 abs}\gapp1.7$), these will usually have low angular diameter distance
 ratios. This manifests itself as 21-cm absorption being more likely
 to be detected in Mg{\sc \,ii} selected absorbers. This is evident in
 the fact that, although the 21-cm line strength is correlated with the Mg{\sc \,ii}
 equivalent width, it is not decisive in whether 21-cm absorption is
 detected or not, with the non-detections spanning a similar range of
 equivalent widths as the detections. Investigating this and other
 correlations further, we
 find:
%\begin{enumerate}

\item The relationship between the 21-cm line strength and the Mg{\sc
\,ii} equivalent width (significant at $\leq1.9\sigma$) is dominated
by the velocity width of the 21-cm line, thus indicating a correlation
between the 21-cm and Mg{\sc \,ii} velocity profiles
($\leq2.8\sigma$).

\item Since the Mg{\sc \,ii} equivalent width is also correlated with
  the metallicity, this would suggest a 21-cm line
  strength--metallicity relation, which we find at a $\leq2.8\sigma$
  significance.

\item Such a correlation has previously been suggested, on
 evolutionary grounds, by \citet{kb04}. However, although the
 metallicity is known to decrease with redshift for larger or more
 homogeneous samples of DLAs, no metallicity--redshift correlation is
 seen for this sample, suggesting that this relationship may be
 non-evolutionary in origin for the 21-cm absorbing DLAs. This is
 confirmed by our finding that the $T_{\rm spin}/{f}$ ratio does not
 appear to exhibit an overall increase with redshift.
\end{enumerate}
Although the relationships between these various parameters are
expected to be complex and intricately interconnected, we suggest
since ${\rm [M/H]}\propto{\rm W}_{\rm r}^{\lambda2796}\propto \Delta
V_{\rm MgII} \propto \Delta V_{\rm 21-cm}$, that the observed 21-cm
line strength--metallicity correlation is a consequence of the
coupling between the velocity structure of the cold neutral (\HI\
21-cm) and the singly ionised (Mg{\sc \,ii}) gas.
%That is, it can be argued that the absorption at 21-cm is due to the same gas which creates the optical absorption lines.

\section*{Acknowledgments}

We would like to thank the anonymous referee for their very helpful
comments, Raffaella Morganti for coordinating the WSRT observations,
Michael Murphy for analysing the SDSS spectra of 1402+044,Martin
Zwaan, Matthew Whiting and Martin Thompson for their advice.  The
Westerbork Synthesis Radio Telescope is operated by the ASTRON
(Netherlands Foundation for Research in Astronomy) with support from
the Netherlands Foundation for Scientific Research NWO.  This research
has made use of the NASA/IPAC Extragalactic Database (NED) which is
operated by the Jet Propulsion Laboratory, California Institute of
Technology, under contract with the National Aeronautics and Space
Administration. This research has also made use of NASA's Astrophysics
Data System Bibliographic Services.

\label{lastpage}

\begin{thebibliography}{}

\bibitem[\protect\citeauthoryear{Akerman, Ellison, Pettini \& Steidel}{Akerman
  et~al.}{2005}]{aeps05}
Akerman C.~J.,  Ellison S.~L.,  Pettini M.,    Steidel C.~C.,  2005, A\&A, 440,
  499

\bibitem[\protect\citeauthoryear{{Bergeron} \& {Boiss\'{e}}}{{Bergeron} \&
  {Boiss\'{e}}}{1991}]{bb91}
{Bergeron} J.,  {Boiss\'{e}} P.,  1991, A\&A, 243, 344

\bibitem[\protect\citeauthoryear{Bouch\'{e}, Murphy, P\'{e}roux, Csabai \&
  Wild}{Bouch\'{e} et~al.}{2006}]{bmp+06}
Bouch\'{e} N.,  Murphy M.~T.,  P\'{e}roux C.,  Csabai I.,    Wild V.,  2006,
  MNRAS, 371, 495

\bibitem[\protect\citeauthoryear{{Briggs}, {Brinks} \& {Wolfe}}{{Briggs}
  et~al.}{1997}]{bbw97}
{Briggs} F.~H.,  {Brinks} E.,    {Wolfe} A.~M.,  1997, AJ, 113, 467

\bibitem[\protect\citeauthoryear{Briggs \& Wolfe}{Briggs \& Wolfe}{1983}]{bw83}
Briggs F.~H.,  Wolfe A.~M.,  1983, ApJ, 268, 76

\bibitem[\protect\citeauthoryear{{Brown} \& {Roberts}}{{Brown} \&
  {Roberts}}{1973}]{br73}
{Brown} R.~L.,  {Roberts} M.~S.,  1973, ApJ, 184, L7

\bibitem[\protect\citeauthoryear{{Carilli}, {Perlman} \& {Stocke}}{{Carilli}
  et~al.}{1992}]{cps92}
{Carilli} C.~L.,  {Perlman} E.~S.,    {Stocke} J.~T.,  1992, ApJ, 400, L13

\bibitem[\protect\citeauthoryear{{Chen}, {Lanzetta}, {Webb} \&
  {Barcons}}{{Chen} et~al.}{1998}]{clwb98}
{Chen} H.,  {Lanzetta} K.~M.,  {Webb} J.~K.,    {Barcons} X.,  1998, ApJ, 498,
  77

\bibitem[\protect\citeauthoryear{Chen \& Lanzetta}{Chen \&
  Lanzetta}{2003}]{cl03}
Chen H.-W.,  Lanzetta K.~M.,  2003, ApJ, 597, 706

\bibitem[\protect\citeauthoryear{{Chengalur} \& {Kanekar}}{{Chengalur} \&
  {Kanekar}}{2000}]{ck00}
{Chengalur} J.~N.,  {Kanekar} N.,  2000, MNRAS, 318, 303

\bibitem[\protect\citeauthoryear{{Cui}, {Bechtold}, {Ge} \& {Meyer}}{{Cui}
  et~al.}{2005}]{cbgm05}
{Cui} J.,  {Bechtold} J.,  {Ge} J.,    {Meyer} D.~M.,  2005, ApJ, 633, 649

\bibitem[\protect\citeauthoryear{Curran, Darling, Bolatto, Whiting, Bignell \&
  Webb}{Curran et~al.}{2007\natexlab{a}}]{cdbw07}
Curran S.~J.,  Darling J.~K.,  Bolatto A.~D.,  Whiting M.~T.,  Bignell C.,
  Webb J.~K.,  2007{\natexlab{a}}, MNRAS, in press (arXiv:0708.1636)

\bibitem[\protect\citeauthoryear{Curran, Murphy, Pihlstr\"{o}m, Webb \&
  Purcell}{Curran et~al.}{2005}]{cmp+03}
Curran S.~J.,  Murphy M.~T.,  Pihlstr\"{o}m Y.~M.,  Webb J.~K.,    Purcell
  C.~R.,  2005, MNRAS, 356, 1509

\bibitem[\protect\citeauthoryear{Curran, Tzanavaris, Murphy, Webb \&
  Pihlstr\"{o}m}{Curran et~al.}{2007\natexlab{b}}]{ctm+07}
Curran S.~J.,  Tzanavaris P.,  Murphy M.~T.,  Webb J.~K.,    Pihlstr\"{o}m
  Y.~M.,  2007{\natexlab{b}}, MNRAS, in press (arXiv:0706.2692)

\bibitem[\protect\citeauthoryear{Curran \& Webb}{Curran \& Webb}{2006}]{cw06}
Curran S.~J.,  Webb J.~K.,  2006, MNRAS, 371, 356

\bibitem[\protect\citeauthoryear{Curran, Webb, Murphy, Bandiera, Corbelli \&
  Flambaum}{Curran et~al.}{2002}]{cwbc01}
Curran S.~J.,  Webb J.~K.,  Murphy M.~T.,  Bandiera R.,  Corbelli E.,
  Flambaum V.~V.,  2002, PASA, 19, 455

\bibitem[\protect\citeauthoryear{Curran, Webb, Murphy \& Carswell}{Curran
  et~al.}{2004}]{cwmc03}
Curran S.~J.,  Webb J.~K.,  Murphy M.~T.,    Carswell R.~F.,  2004, MNRAS, 351,
  L24

\bibitem[\protect\citeauthoryear{Curran, Whiting, Murphy, Webb, Longmore,
  Pihlstr\"{o}m, Athreya \& Blake}{Curran et~al.}{2006}]{cwm+06}
Curran S.~J.,  Whiting M.,  Murphy M.~T.,  Webb J.~K.,  Longmore S.~N.,
  Pihlstr\"{o}m Y.~M.,  Athreya R.,    Blake C.,  2006, MNRAS, 371, 431

\bibitem[\protect\citeauthoryear{{Dallacasa}, {Bondi}, {Alef} \&
  {Mantovani}}{{Dallacasa} et~al.}{1998}]{dbam98}
{Dallacasa} D.,  {Bondi} M.,  {Alef} W.,    {Mantovani} F.,  1998, A\&AS, 129,
  219

\bibitem[\protect\citeauthoryear{{de Bruyn}, {O'Dea} \& {Baum}}{{de Bruyn}
  et~al.}{1996}]{dob96}
{de Bruyn} A.~G.,  {O'Dea} C.~P.,    {Baum} S.~A.,  1996, A\&A, 305, 450

\bibitem[\protect\citeauthoryear{{Ellison}}{{Ellison}}{2006}]{ell06}
{Ellison} S.~L.,  2006, MNRAS, 368, 335

\bibitem[\protect\citeauthoryear{{Ellison}, {Pettini}, {Steidel} \&
  {Shapley}}{{Ellison} et~al.}{2001}]{epss01}
{Ellison} S.~L.,  {Pettini} M.,  {Steidel} C.~C.,    {Shapley} A.~E.,  2001,
  ApJ, 549, 770

\bibitem[\protect\citeauthoryear{{Fox}, {Ledoux}, {Petitjean} \&
  {Srianand}}{{Fox} et~al.}{2007}]{flps07}
{Fox} A.~J.,  {Ledoux} C.,  {Petitjean} P.,    {Srianand} R.,  2007, A\&A,
  accepted (arXiv:0707.4065)

\bibitem[\protect\citeauthoryear{{Ge}, {Bechtold} \& {Kulkarni}}{{Ge}
  et~al.}{2001}]{gbk01}
{Ge} J.,  {Bechtold} J.,    {Kulkarni} V.~P.,  2001, ApJ, 547, L1

\bibitem[\protect\citeauthoryear{Gupta, Srianand, Petitjean, Khare, Saikia \&
  York}{Gupta et~al.}{2007}]{gsp+06}
Gupta N.,  Srianand R.,  Petitjean P.,  Khare P.,  Saikia D.~J.,    York D.~G.,
   2007, ApJ, 654, L111

\bibitem[\protect\citeauthoryear{Haehnelt, Steinmetz \& Rauch}{Haehnelt
  et~al.}{1998}]{hsr98}
Haehnelt M.~G.,  Steinmetz M.,    Rauch M.,  1998, ApJ, 495, 647

\bibitem[\protect\citeauthoryear{Kanekar \& Briggs}{Kanekar \&
  Briggs}{2004}]{kb04}
Kanekar N.,  Briggs F.~H.,  2004, Science with the Square Kilometer Array, New
  Astronomy Reviews.
Elsevier, Amsterdam, pp 1259--1270

\bibitem[\protect\citeauthoryear{Kanekar \& Chengalur}{Kanekar \&
  Chengalur}{2001}]{kc01a}
Kanekar N.,  Chengalur J.~N.,  2001, A\&A, 369, 42

\bibitem[\protect\citeauthoryear{Kanekar \& Chengalur}{Kanekar \&
  Chengalur}{2003}]{kc02}
Kanekar N.,  Chengalur J.~N.,  2003, A\&A, 399, 857

\bibitem[\protect\citeauthoryear{{Kanekar}, {Chengalur} \& {Lane}}{{Kanekar}
  et~al.}{2007}]{kcl06}
{Kanekar} N.,  {Chengalur} J.~N.,    {Lane} W.~M.,  2007, MNRAS, 375, 1528

\bibitem[\protect\citeauthoryear{{Kanekar}, {Subrahmanyan}, {Ellison}, {Lane}
  \& {Chengalur}}{{Kanekar} et~al.}{2006}]{kse+06}
{Kanekar} N.,  {Subrahmanyan} R.,  {Ellison} S.~L.,  {Lane} W.,    {Chengalur}
  J.~N.,  2006, MNRAS, 370, L46

\bibitem[\protect\citeauthoryear{{Khare}, {Kulkarni}, {P{\'e}roux}, {York},
  {Lauroesch} \& {Meiring}}{{Khare} et~al.}{2007}]{kkp+07}
{Khare} P.,  {Kulkarni} V.~P.,  {P{\'e}roux} C.,  {York} D.~G.,  {Lauroesch}
  J.~T.,    {Meiring} J.~D.,  2007, A\&A, 464, 487

\bibitem[\protect\citeauthoryear{Kulkarni \& Fall}{Kulkarni \&
  Fall}{2002}]{kf02}
Kulkarni V.~P.,  Fall S.~M.,  2002, ApJ, 580, 732

\bibitem[\protect\citeauthoryear{Lane}{Lane}{2000}]{lan00}
Lane W.~M.,  2000, PhD thesis, University of Groningen

\bibitem[\protect\citeauthoryear{{Lane}, {Briggs} \& {Smette}}{{Lane}
  et~al.}{2000}]{lbs00}
{Lane} W.~M.,  {Briggs} F.~H.,    {Smette} A.,  2000, ApJ, 532, 146

\bibitem[\protect\citeauthoryear{{Le Brun}, Bergeron, Boiss\'{e} \&
  Deharveng}{{Le Brun} et~al.}{1997}]{lbbd97}
{Le Brun} V.,  Bergeron J.,  Boiss\'{e} P.,    Deharveng J.~M.,  1997, A\&A,
  321, 733

\bibitem[\protect\citeauthoryear{{Le Brun}, {Viton} \& {Milliard}}{{Le Brun}
  et~al.}{1998}]{lvm98}
{Le Brun} V.,  {Viton} M.,    {Milliard} B.,  1998, A\&A, 340, 381

\bibitem[\protect\citeauthoryear{{Ledoux}, {Petitjean}, {Fynbo} \& {M\o ller}
  P.and~{Srianand}}{{Ledoux} et~al.}{2006}]{lpf+06}
{Ledoux} C.,  {Petitjean} P.,  {Fynbo} J.~U.,    {M\o ller} P.and~{Srianand}
  R.,  2006, A\&A, 457, 71

\bibitem[\protect\citeauthoryear{Ledoux, Petitjean \& Srianand}{Ledoux
  et~al.}{2003}]{lps03}
Ledoux C.,  Petitjean P.,    Srianand R.,  2003, MNRAS, 346, 209

\bibitem[\protect\citeauthoryear{Ledoux, Petitjean \& Srianand}{Ledoux
  et~al.}{2006}]{lps06}
Ledoux C.,  Petitjean P.,    Srianand R.,  2006, ApJ, 640, L25

\bibitem[\protect\citeauthoryear{Ledoux, Srianand \& Petitjean}{Ledoux
  et~al.}{2002}]{lsp02}
Ledoux C.,  Srianand R.,    Petitjean P.,  2002, A\&A, 392, 781

\bibitem[\protect\citeauthoryear{Levshakov, Dessauges-Zavadsky, D'Odorico \&
  Molaro}{Levshakov et~al.}{2002}]{lddm01}
Levshakov S.~A.,  Dessauges-Zavadsky M.,  D'Odorico S.,    Molaro P.,  2002,
  ApJ, 565, 696

\bibitem[\protect\citeauthoryear{{Levshakov}, {Molaro}, {Centuri\'{o}n},
  {D'Odorico}, {Bonifacio} \& {Vladilo}}{{Levshakov} et~al.}{2000}]{lmc+00a}
{Levshakov} S.~A.,  {Molaro} P.,  {Centuri\'{o}n} M.,  {D'Odorico} S.,
  {Bonifacio} P.,    {Vladilo} G.,  2000, A\&A, 361, 803

\bibitem[\protect\citeauthoryear{{Levshakov} \& {Varshalovich}}{{Levshakov} \&
  {Varshalovich}}{1985}]{lv85}
{Levshakov} S.~A.,  {Varshalovich} D.~A.,  1985, MNRAS, 212, 517

\bibitem[\protect\citeauthoryear{{Lopez} \& {Ellison}}{{Lopez} \&
  {Ellison}}{2003}]{le03}
{Lopez} S.,  {Ellison} S.~L.,  2003, A\&A, 403, 573

\bibitem[\protect\citeauthoryear{{Lu}, {Sargent}, {Barlow}, {Churchill} \&
  {Vogt}}{{Lu} et~al.}{1996}]{lsb+96}
{Lu} L.,  {Sargent} W. L.~W.,  {Barlow} T.~A.,  {Churchill} C.~W.,    {Vogt}
  S.~S.,  1996, ApJS, 107, 475

\bibitem[\protect\citeauthoryear{{Meiring}, {Lauroesch}, {Kulkarni},
  {P{\'e}roux}, {Khare}, {York} \& {Crotts}}{{Meiring} et~al.}{2007}]{mlk+07}
{Meiring} J.~D.,  {Lauroesch} J.~T.,  {Kulkarni} V.~P.,  {P{\'e}roux} C.,
  {Khare} P.,  {York} D.~G.,    {Crotts} A.~P.~S.,  2007, MNRAS, 376, 557

\bibitem[\protect\citeauthoryear{Murphy, Curran \& Webb}{Murphy
  et~al.}{2004}]{mcw04}
Murphy M.~T.,  Curran S.~J.,    Webb J.~K.,  2004, in Duc P.-A.,  Braine J.,
  Brinks E.,  eds, Recycling Intergalactic and Interstellar Matter, IAU
  Symposium No. 217 H$_2$-bearing damped lyman-{$\alpha$} systems as tracers of
  cosmological chemical evolution.
ASP Conf. Ser., San Francisco, p.~252

\bibitem[\protect\citeauthoryear{Murphy, Curran, Webb, M\'{e}nager \&
  Zych}{Murphy et~al.}{2007}]{mcw+07}
Murphy M.~T.,  Curran S.~J.,  Webb J.~K.,  M\'{e}nager H.,    Zych B.~J.,
  2007, MNRAS, 376, 673

\bibitem[\protect\citeauthoryear{{Murphy} \& {Liske}}{{Murphy} \&
  {Liske}}{2004}]{ml04}
{Murphy} M.~T.,  {Liske} J.,  2004, MNRAS, 354, L31

\bibitem[\protect\citeauthoryear{{Nestor}, {Rao}, {Turnshek} \& {Vanden
  Berk}}{{Nestor} et~al.}{2003}]{nrtv03}
{Nestor} D.~B.,  {Rao} S.~M.,  {Turnshek} D.~A.,    {Vanden Berk} D.,  2003,
  ApJ, 595, L5

\bibitem[\protect\citeauthoryear{Noterdaeme, Ledoux, , Petitjean, Petit,
  Srianand \& Smette}{Noterdaeme et~al.}{2007}]{nlp+07}
Noterdaeme P.,  Ledoux C.,   Petitjean P.,  Petit F.~L.,  Srianand R.,
  Smette A.,  2007, A\&A, accepted (arXiv:0707.4479)

\bibitem[\protect\citeauthoryear{{P{\'e}roux}, {Dessauges-Zavadsky},
  {D'Odorico}, {Kim} \& {McMahon}}{{P{\'e}roux} et~al.}{2003}]{pdd+03}
{P{\'e}roux} C.,  {Dessauges-Zavadsky} M.,  {D'Odorico} S.,  {Kim} T.-S.,
  {McMahon} R.~G.,  2003, MNRAS, 345, 480

\bibitem[\protect\citeauthoryear{{Petitjean}, {Ledoux}, {Noterdaeme} \&
  {Srianand}}{{Petitjean} et~al.}{2006}]{plns06}
{Petitjean} P.,  {Ledoux} C.,  {Noterdaeme} P.,    {Srianand} R.,  2006, A\&A,
  456, L9

\bibitem[\protect\citeauthoryear{{Petitjean}, {Srianand} \&
  {Ledoux}}{{Petitjean} et~al.}{2002}]{psl02}
{Petitjean} P.,  {Srianand} R.,    {Ledoux} C.,  2002, MNRAS, 332, 383

\bibitem[\protect\citeauthoryear{{Pettini}, {King}, {Smith} \&
  {Hunstead}}{{Pettini} et~al.}{1995}]{pksh95}
{Pettini} M.,  {King} D.~L.,  {Smith} L.~J.,    {Hunstead} R.~W.,  1995, in
  Meylan G.,  ed., QSO Absorption Lines {The Chemical Evolution of Damped
  Lyman-Alpha Galaxies}.
Springer-Verlag, Berlin, p.~71

\bibitem[\protect\citeauthoryear{{Prochaska}, {Chen}, {Wolfe},
  {Dessauges-Zavadsky} \& {Bloom}}{{Prochaska} et~al.}{2007}]{pcw+07}
{Prochaska} J.~X.,  {Chen} H.-W.,  {Wolfe} A.~M.,  {Dessauges-Zavadsky} M.,
  {Bloom} J.~S.,  2007, ApJ, submitted (astro-ph/0703701)

\bibitem[\protect\citeauthoryear{Prochaska, Gawiser, Wolfe, Castro \&
  Djorgovski}{Prochaska et~al.}{2003}]{pgw+03}
Prochaska J.~X.,  Gawiser E.,  Wolfe A.~M.,  Castro S.,    Djorgovski S.~G.,
  2003, ApJ, 595, L9

\bibitem[\protect\citeauthoryear{Prochaska \& Herbert-Fort}{Prochaska \&
  Herbert-Fort}{2004}]{ph04}
Prochaska J.~X.,  Herbert-Fort S.,  2004, PASP, 116, 622

\bibitem[\protect\citeauthoryear{Prochaska, Herbert-Fort \& Wolfe}{Prochaska
  et~al.}{2005}]{phw05}
Prochaska J.~X.,  Herbert-Fort S.,    Wolfe A.~M.,  2005, ApJ, 635, 123

\bibitem[\protect\citeauthoryear{{Prochaska} \& {Wolfe}}{{Prochaska} \&
  {Wolfe}}{1997}]{pw97}
{Prochaska} J.~X.,  {Wolfe} A.~M.,  1997, ApJ, 487, 73

\bibitem[\protect\citeauthoryear{Rao, Nestor, Turnshek, Lane, Monier \&
  Bergeron}{Rao et~al.}{2003}]{rnt+03}
Rao S.,  Nestor D.~B.,  Turnshek D.,  Lane W.~M.,  Monier E.~M.,    Bergeron
  J.,  2003, ApJ, 595, 94

\bibitem[\protect\citeauthoryear{{Rao} \& {Turnshek}}{{Rao} \&
  {Turnshek}}{2000}]{rt00}
{Rao} S.~M.,  {Turnshek} D.~A.,  2000, ApJS, 130, 1

\bibitem[\protect\citeauthoryear{Reimers, Baade, Quast \& Levshakov}{Reimers
  et~al.}{2003}]{rbql03}
Reimers D.,  Baade R.,  Quast R.,    Levshakov S.~A.,  2003, A\&A, 410, 785

\bibitem[\protect\citeauthoryear{Ryan-Weber, Webster \&
  Staveley-Smith}{Ryan-Weber et~al.}{2003}]{rws03}
Ryan-Weber E.~V.,  Webster R.~L.,    Staveley-Smith L.,  2003, MNRAS, 343, 1195

\bibitem[\protect\citeauthoryear{{Srianand}, {Gupta} \& {Petitjean}}{{Srianand}
  et~al.}{2007}]{sgp06}
{Srianand} R.,  {Gupta} N.,    {Petitjean} P.,  2007, MNRAS, 375, 584

\bibitem[\protect\citeauthoryear{{Steidel}, {Dickinson}, {Meyer}, {Adelberger}
  \& {Sembach}}{{Steidel} et~al.}{1997}]{sdm+97}
{Steidel} C.~C.,  {Dickinson} M.,  {Meyer} D.~M.,  {Adelberger} K.~L.,
  {Sembach} K.~R.,  1997, ApJ, 480, 568

\bibitem[\protect\citeauthoryear{{Turnshek}, {Rao}, {Nestor}, {Belfort-Mihalyi}
  \& {Quider}}{{Turnshek} et~al.}{2005}]{trn+05}
{Turnshek} D.~A.,  {Rao} S.~M.,  {Nestor} D.~B.,  {Belfort-Mihalyi} M.,
  {Quider} A.,  2005, in Williams P.~R.,  Shu C.,   M\'{e}nard B.,  eds,
  Probing Galaxies through Quasar Absorption Lines, Proceedings IAU Colloquium No.
  199 (astro-ph/0506701)

\bibitem[\protect\citeauthoryear{{Turnshek}, {Wolfe}, {Lanzetta}, {Briggs},
  {Cohen}, {Foltz}, {Smith} \& {Wilkes}}{{Turnshek} et~al.}{1989}]{twl+89}
{Turnshek} D.~A.,  {Wolfe} A.~M.,  {Lanzetta} K.~M.,  {Briggs} F.~H.,  {Cohen}
  R.~D.,  {Foltz} C.~B.,  {Smith} H.~E.,    {Wilkes} B.~J.,  1989, ApJ, 344,
  567

\bibitem[\protect\citeauthoryear{Tzanavaris, Murphy, Webb, Flambaum \&
  Curran}{Tzanavaris et~al.}{2007}]{tmw+06}
Tzanavaris P.,  Murphy M.~T.,  Webb J.~K.,  Flambaum V.~V.,    Curran S.~J.,
  2007, MNRAS, 374, 634

\bibitem[\protect\citeauthoryear{{Vladilo}, {Bonifacio}, {Centuri\'{o}n} \&
  {Molaro}}{{Vladilo} et~al.}{2000}]{vbcm00}
{Vladilo} G.,  {Bonifacio} P.,  {Centuri\'{o}n} M.,    {Molaro} P.,  2000, ApJ,
  543, 24

\bibitem[\protect\citeauthoryear{Wolfe, Briggs \& Jauncey}{Wolfe
  et~al.}{1981}]{wbj81}
Wolfe A.~M.,  Briggs F.~H.,    Jauncey D.~L.,  1981, ApJ, 248, 460

\bibitem[\protect\citeauthoryear{{Wolfe}, {Briggs}, {Turnshek}, {Davis},
  {Smith} \& {Cohen}}{{Wolfe} et~al.}{1985}]{wbt+85}
{Wolfe} A.~M.,  {Briggs} F.~H.,  {Turnshek} D.~A.,  {Davis} M.~M.,  {Smith}
  H.~E.,    {Cohen} R.~D.,  1985, ApJ, 294, L67

\bibitem[\protect\citeauthoryear{{Wolfe}, {Lanzetta}, {Foltz} \&
  {Chaffee}}{{Wolfe} et~al.}{1995}]{wlfc95}
{Wolfe} A.~M.,  {Lanzetta} K.~M.,  {Foltz} C.~B.,    {Chaffee} F.~H.,  1995,
  ApJ, 454, 698

\bibitem[\protect\citeauthoryear{{Wolfe} \& {Prochaska}}{{Wolfe} \&
  {Prochaska}}{1998}]{wp98}
{Wolfe} A.~M.,  {Prochaska} J.~X.,  1998, ApJ, 494, L15

\bibitem[\protect\citeauthoryear{{Wolfe} \& {Prochaska}}{{Wolfe} \&
  {Prochaska}}{2000}]{wp00}
{Wolfe} A.~M.,  {Prochaska} J.~X.,  2000, ApJ, 545, 591

\bibitem[\protect\citeauthoryear{{Wolfe}, {Turnshek}, {Smith} \&
  {Cohen}}{{Wolfe} et~al.}{1986}]{wtsc86}
{Wolfe} A.~M.,  {Turnshek} D.~A.,  {Smith} H.~E.,    {Cohen} R.~D.,  1986,
  ApJS, 61, 249

\bibitem[\protect\citeauthoryear{{Wolfire}, {Hollenbach}, {McKee}, {Tielens} \&
  {Bakes}}{{Wolfire} et~al.}{1995}]{whm+95}
{Wolfire} M.~G.,  {Hollenbach} D.,  {McKee} C.~F.,  {Tielens} A.~G.~G.~M.,
  {Bakes} E.~L.~O.,  1995, ApJ, 443, 152

\end{thebibliography}
\end{document}